\documentclass[
    ,final            
  ,numberedheadings 
  ]
  {aipproc}

\layoutstyle{6x9}

\newcommand{\bs}{\begin{slide}}
\newcommand{\es}{\end{slide}}
\newcommand{\be}{\begin{equation}}
\newcommand{\ee}{\end{equation}}
\newcommand{\bea}{\begin{eqnarray}}
\newcommand{\eea}{\end{eqnarray}}

\newcommand{\la}{\left\langle}
\newcommand{\ra}{\right\rangle}
\newcommand{\lc}{\left[}
\newcommand{\rc}{\right]}
\newcommand{\lp}{\left(}
\newcommand{\rp}{\right)}

\newcommand{\aq}{\alpha_s\lp Q^2\rp}
\newcommand{\aqq}{\alpha_s\lp Q_0^2\rp}
\newcommand{\bc}{\begin{center}}
\newcommand{\ec}{\end{center}}

\newcommand{\bi}{\begin{itemize}}
\newcommand{\ei}{\end{itemize}}
\def\epm#1#2{\hbox{${\lower1pt\hbox{$\scriptstyle +~#1$}}
\atop {\raise1pt\hbox{$\scriptstyle -~#2$}}$}}
\newcommand{\dat}{\mathrm{dat}}
\newcommand{\art}{\mathrm{art}}
\newcommand{\net}{\mathrm{net}}
\newcommand{\rep}{\mathrm{rep}}
\newcommand{\rmexp}{\mathrm{exp}}

\begin{document}

\title{The neural network approach to parton fitting\footnote{
Talk given at Deep Inelastic Scattering 2005 Workshop (Madison) by 
J. R., on behalf of the NNPDF Collaboration.}}


\author{The NNPDF Collaboration: Joan Rojo}{
  address={  Departament d'Estructura i Constituents de la Mat\`eria, 
Universitat de Barcelona}
}

\author{Luigi Del Debbio}{
  address={Theory Division, CERN}
}

\author{Stefano Forte}{
  address={ Dipartimento di Fisica, Universit\`a di Milano and 
INFN, Sezione di Milano}
}

\author{Jos\'e I. Latorre}{
  address={ Departament d'Estructura i Constituents de la Mat\`eria, 
Universitat de Barcelona}
}

\author{Andrea Piccione}{
  address={Dipartimento di Fisica Teorica, Universit\`a di Torino, and
INFN Sezione di Torino }
}

\begin{abstract}
We introduce the neural network approach to global fits
of parton distribution functions. First we review previous
work on unbiased parametrizations of deep-inelastic 
structure functions with 
faithful estimation of their uncertainties, and then we
summarize the current status of neural network parton distribution fits.
\end{abstract}

\maketitle


\section{Introduction}

The requirements of precision physics at hadron colliders have recently led
to a rapid improvement in the techniques for the determination of 
parton distributions of the nucleon \cite{tung}. 
Specifically it is now mandatory to 
determine accurately the uncertainty on these quantities. The
main difficulty
 is that one is trying to determine the uncertainty on a function,
that is, a probability measure in a space of functions, and to extract it from
a finite set of experimental data, a problem which is mathematically ill-posed 

The shortcomings of the standard approach to global parton fits 
 are well
known: the bias introduced by choosing fixed functional forms to
parametrize the parton distributions
(also known as {\it model dependence}), the problems to assess
faithfully their uncertainties, the
combination of inconsistent experiments, and the lack
of general, process-independent error propagation techniques.
Although the problem of quantifying the uncertainties in pdfs has seen
a huge progress since its paramount importance was raised some
years ago, until now no unambiguous conclusions have been obtained.

Here we present a novel strategy to address the problem
of constructing unbiased parametrizations of parton distributions
with a faithful estimation of their uncertainties, based on 
a combination of two techniques: Monte Carlo methods and neural networks.
First we review recent work on the related problem 
of the construction of bias-free parametrizations of structure
functions from experimental data, and then we turn to the
application of our strategy to parton distributions.

\section{Structure functions}
The strategy presented in \cite{f2nn,nnpdf} to address to problem
of parametrizing deep-inelastic structure functions $F(x,Q^2)$ 
is a combination of two techniques: 
first we construct a Monte Carlo sampling of the experimental data 
(generating artificial data replicas), and then
we train neural networks in each data replica, to
construct a probability measure in the space of structure functions
$\mathcal{P}\lc F(x,Q^2)\rc$. The probability measure constructed
in this way
contains all information from experimental data, including correlations,
with the only assumption of smoothness. Expectation values and moments over
this probability measure are then evaluated as averages over
the trained network sample,
\be
\label{probmeas}
\la \mathcal{F}\lc F(x,Q^2)\rc\ra=\int\mathcal{D}F
\mathcal{P}\lc F(x,Q^2)\rc
\mathcal{F}\lc F(x,Q^2)\rc=\frac{1}{N_{\rep}}
\sum_{k=1}^{N_{\rep}}\mathcal{F}\lp F^{(\net)(k)}(x,Q^2)\rp \ .
\ee
where $\mathcal{F}\lc F\rc$ is an arbitrary function of $F(x,Q^2)$.

The first step is the Monte Carlo sampling of experimental data, 
generating $N_{\rep}$ replicas of the original $N_{\dat}$ experimental data,
\be
F_i^{(\art)(k)} =\lp 1+r_N^{(k)}\sigma_N\rp\lc F_i^{(\rmexp)}+r_i^{s,(k)}
\sigma^{stat}_i+\sum_{l=1}^{N_{sys}}r^{l,(k)}\sigma_i^{sys,l} \rc, 
\qquad i=1,\ldots,N_{\dat} \ ,
\ee
where $r$ are gaussian random numbers with the same correlation
as the respective uncertainties, and $\sigma^{stat},\sigma^{sys},
\sigma_{N}$ are the statistical, systematic and normalization
errors.
The number of replicas $N_{\rep}$ has to be large enough so that the
replica sample
 reproduces central values, errors and correlations
of the experimental data.

\begin{figure}
  \includegraphics[height=.20\textheight]{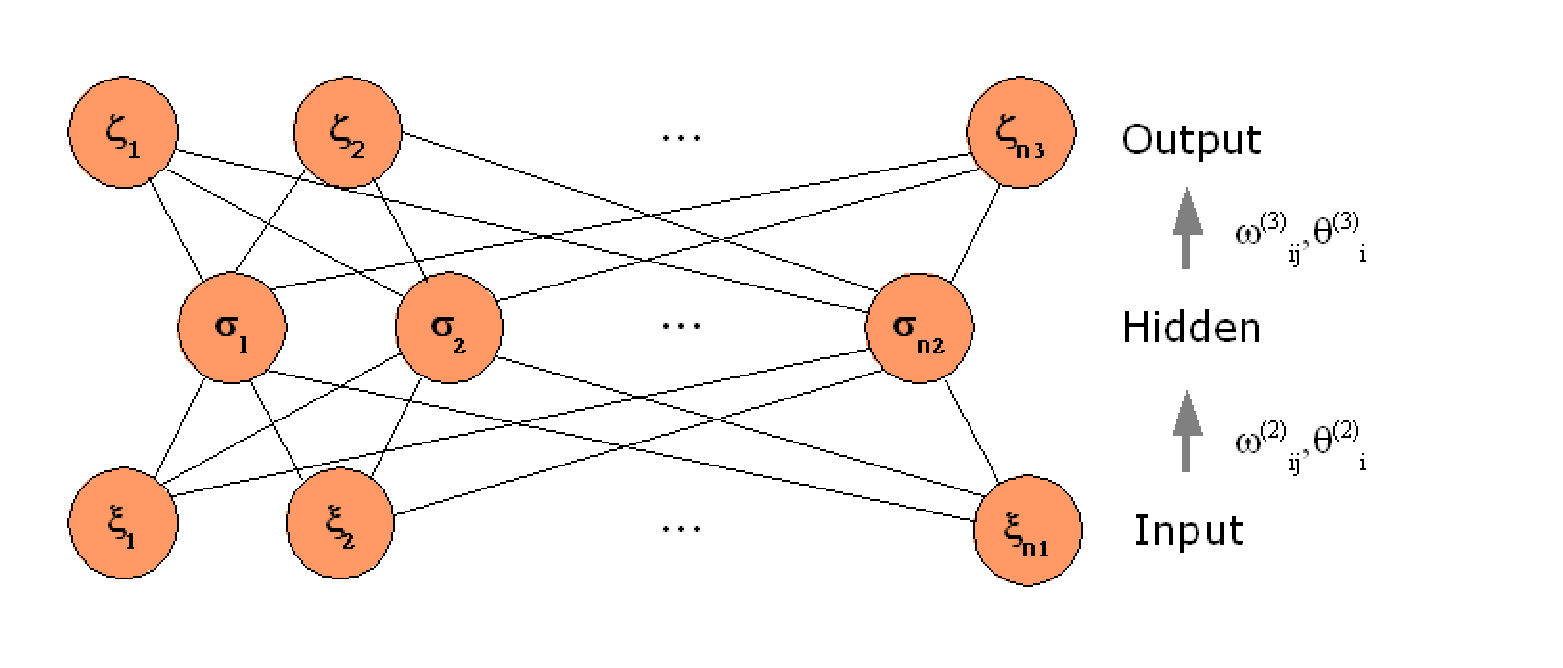}
  \includegraphics[height=.20\textheight]{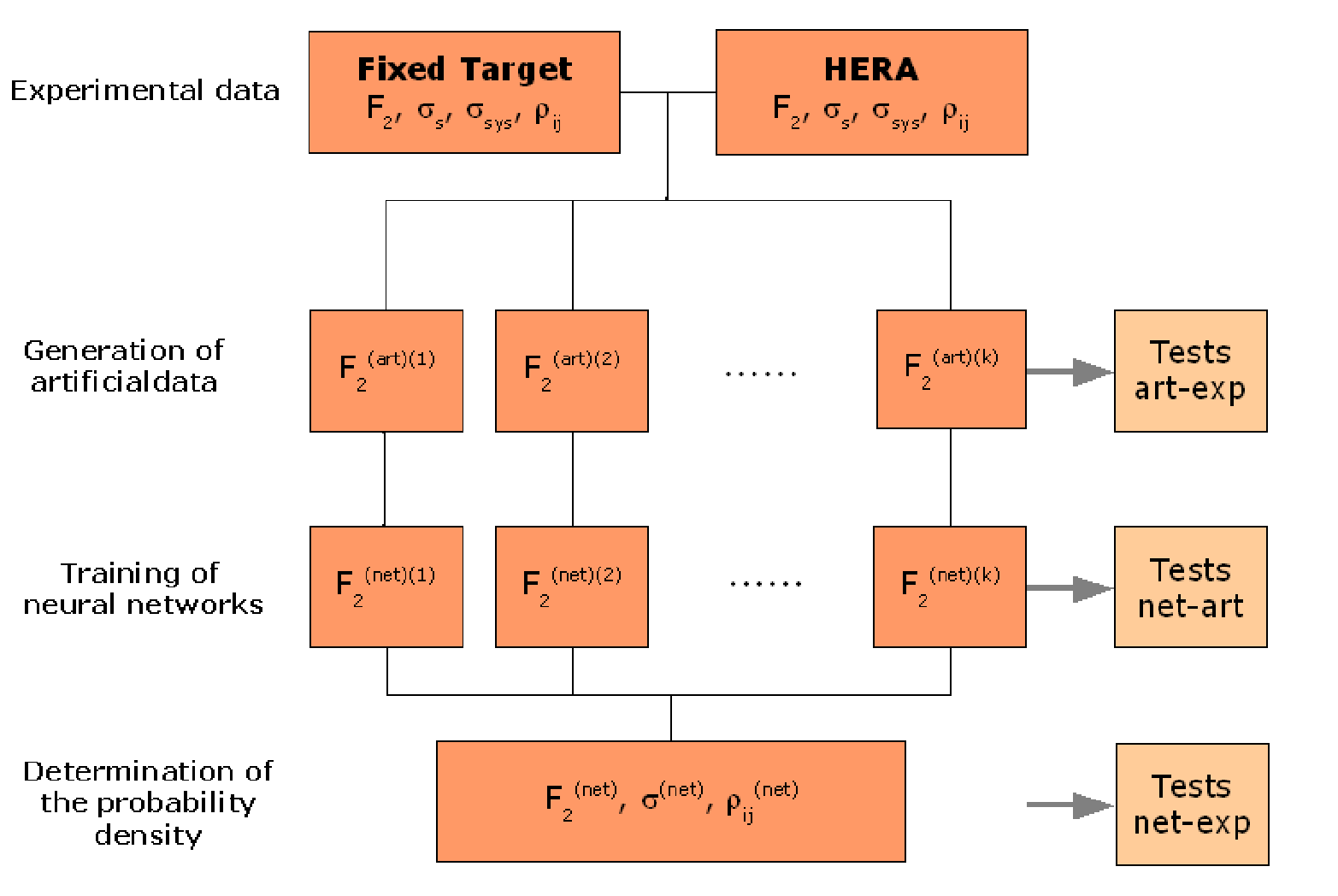}
  \caption{Left: a neural network. Right: Summary of the NNPDF approach.}
\label{nn}
\end{figure}

The second step consists on training a neural network on each of the data
replicas. A neural network \cite{nn} (see fig. \ref{nn}) is a highly
nonlinear mapping between input and output patterns as a function
of its parameters (the so-called {\it weights} $\omega_{ij}^{(l)}$and 
{\it thresholds} $\theta_i^{(l)}$). Neural networks
are specially suitable to parametrize parton distributions since
they are unbiased, robust approximants and interpolate between
data points with the only assumption of smoothness. The neural network 
training consist on the minimization for each replica of the
$\chi^2$ defined with the inverse of the 
experimental covariance matrix,
\be
{\chi^2}^{(k)}=\frac{1}{N_{\dat}}\sum_{i,j=1}^{N_{\dat}}\lp
F_i^{(\art)(k)}-F_i^{(\net)(k)}\rp\mathrm{cov}^{-1}_{ij}
\lp F_j^{(\art)(k)}-F_j^{(\net)(k)}\rp \ .
\ee
Our minimization strategy is based on
Genetic Algorithms \cite{rojo04}, which are specially suited
for finding global minima in highly nonlinear 
minimization problems.


The set of trained nets, once is validated through suitable statistical
estimators, becomes the sought-for probability measure 
$\mathcal{P}\lc F(x,Q^2)\rc$ in the space of structure functions. 
Now observables with
errors and correlations can be computed from averages over this
probability measure, using eq. (\ref{probmeas}). 
For example, the average and error of a
structure function $F(x,Q^2)$ at arbitrary $(x,Q^2)$ can be
computed as
\be
\la F(x,Q^2) \ra =\frac{1}{N_{\rep}}\sum_{k=1}^{N_{\rep}}
F^{(\net)(k)}(x,Q^2), \quad
\sigma(x,Q^2)=\sqrt{\la  F(x,Q^2)^2 \ra-\la F(x,Q^2) \ra^2} \ .
\ee
Our strategy is summarized in fig. \ref{nn}.
In fig. \ref{f2nn} we
show our results\footnote{
The source code, driver program and graphical web interface for
our structure function fits is available at {
\bf http://sophia.ecm.ub.es/f2neural}.} for the proton structure function
$F_2^p(x,Q^2)$, both in our original fit without HERA data
\cite{f2nn} and in the latest fit \cite{nnpdf} including HERA data.

\section{Parton distributions}
The strategy presented in the above section can be used to parametrize
parton distributions, provided one now takes into 
account Altarelli-Parisi QCD evolution. Therefore we need to define
a suitable evolution formalism, and we will consider for the sake
of simplicity nonsinglet parton evolution. Since complex
neural networks are not allowed, we must use the convolution theorem to evolve
parton distributions in $x-$space using the inverse $\Gamma(x)$
of the Mellin space evolution factor $\Gamma(N)$, defined as
\be
q(N,Q^2)=q(N,Q_0^2) 
\Gamma\lp N,\aq,\aqq\rp \ ,
\ee
\be
\label{gammax}
\Gamma\lp x,\aq,\aqq \rp\equiv\frac{1}{2\pi i}\int_{c-i\infty}^{c+i\infty}
dN~x^{-N}\Gamma\lp N,\aq,\aqq \rp \ .
\ee
The only subtlety is that eq. (\ref{gammax}) defines a distribution,
which must therefore be
regulated at $x=1$, yielding the final evolution equation,
\be
q(x,Q^2)= q(x,Q_0^2)\int_x^1 dy~\Gamma(y) +\int_x^1\frac{dy}{y}
\Gamma(y)\lp q\lp\frac{x}{y},Q_0^2\rp -yq(x,Q_0^2)\rp \ ,
\ee
where in the above equation $q(x,Q_0^2)$ is parametrized
using a neural network.
At higher orders in perturbation theory coefficient functions $C(N)$
are introduced through a modified evolution factor, $\tilde{\Gamma}(N)\equiv
 \Gamma(N) C(N)$.
We have benchmarked our evolution code with the
Les Houches benchmark tables \cite{lh}.
The evolution factor $\Gamma(x)$ and its integral are
computed and interpolated before
the neural network training in order to have a faster fitting
procedure.

\begin{figure}
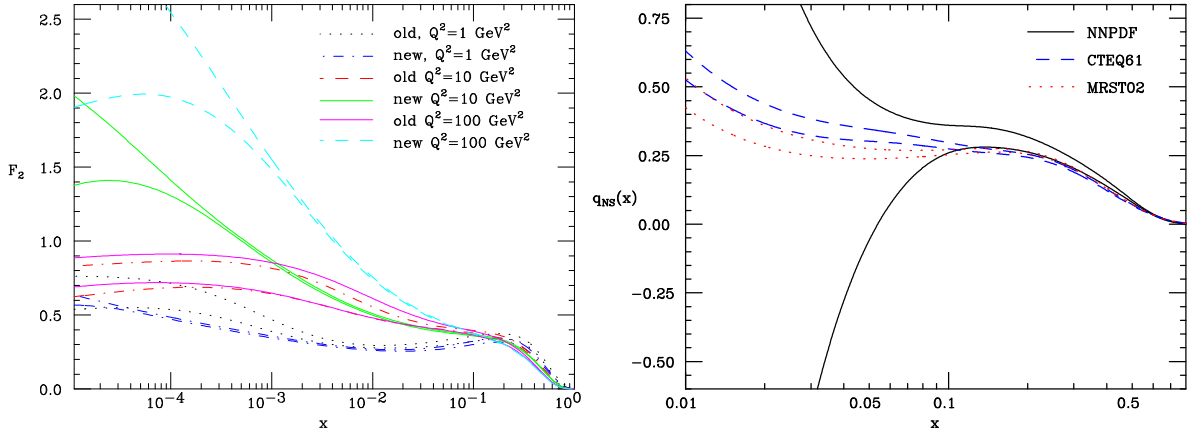

  \includegraphics[height=.26\textheight]{f2_x_log.ps}
 \includegraphics[height=.26\textheight]{qns_nnpdf.ps}
  \caption{Left: The proton structure function, both the
old version \cite{f2nn} without HERA data and the new
version \cite{nnpdf} which included HERA data. Right: Preliminary
results on the nonsinglet parton distribution $q_{NS}(x,Q_0^2)$ 
compared with other global fits, for $Q_0^2=2~\mathrm{GeV}^2$}
\label{f2nn}
\end{figure}

As a first application of our method, we extract the nonsinglet
parton distribution $q_{NS}(x,Q^2_0)=\lp u+\bar{u}-d-\bar{d}\rp(x,Q^2_0)$
from the nonsinglet structure function $F_2^{NS}(x,Q^2)$ measured by the
NMC \cite{nmc} and BCDMS \cite{bcdmsp,bcdmsd} collaborations. 
The preliminary results of a NLO 
fit with fully correlated uncertainties can be 
seen in fig. \ref{f2nn}. Our result is consistent within the
error bands with the results from other global fits 
\cite{mrst01e,cteq61,ale02} in almost all the
range of Bjorken-$x$. It is clear that 
the large uncertainties at small $x$ do not allow,
within the current experimental data, to determine if
$q_{NS}(x,Q^2)$ grows at small $x$, as supported
by different theoretical arguments as well as 
by other global parton fits.
Only additional nonsinglet structure function data at small $x$
can settle this issue.

Summarizing, we have described a general technique to parametrize
experimental data in an bias-free way with a faithful
estimation of their uncertainties, which has been successfully applied to
structure functions and that now is being implemented in the context of  
global parton distribution fits.







\begin{thebibliography}{12}
\expandafter\ifx\csname natexlab\endcsname\relax\def\natexlab#1{#1}\fi
\providecommand{\enquote}[1]{``#1''}
\expandafter\ifx\csname url\endcsname\relax
  \def\url#1{\texttt{#1}}\fi
\expandafter\ifx\csname urlprefix\endcsname\relax\def\urlprefix{URL }\fi
\providecommand{\eprint}[2][]{\url{#2}}

\bibitem[Tung(2005)]{tung}
W.-K. Tung, \emph{AIP Conf. Proc.}, \textbf{753}, 15--29 (2005),
  \eprint{hep-ph/0410139}.

\bibitem[Forte et~al.(2002)]{f2nn}
S.~Forte, L.~Garrido, J.~I. Latorre, and A.~Piccione, \emph{JHEP}, \textbf{05},
  062 (2002), \eprint{hep-ph/0204232}.

\bibitem[Del~Debbio et~al.(2004)]{nnpdf}
L.~Del~Debbio, S.~Forte, J.~I. Latorre, A.~Piccione, and J.~Rojo (2004),
  \eprint{hep-ph/0501067}.

\bibitem[Stimpfl-Abele and Garrido(1991)]{nn}
G.~Stimpfl-Abele, and L.~Garrido, \emph{Comput. Phys. Commun.}, \textbf{64},
  46--56 (1991).

\bibitem[Rojo and Latorre(2004)]{rojo04}
J.~Rojo, and J.~I. Latorre, \emph{JHEP}, \textbf{01}, 055 (2004),
  \eprint{hep-ph/0401047}.

\bibitem[Giele et~al.(2002)]{lh}
W.~Giele, et~al. (2002), \eprint{hep-ph/0204316}.

\bibitem[Arneodo et~al.(1997)]{nmc}
M.~Arneodo, et~al., \emph{Nucl. Phys.}, \textbf{B483}, 3--43 (1997),
  \eprint{hep-ph/9610231}.

\bibitem[Benvenuti et~al.(1989)]{bcdmsp}
A.~C. Benvenuti, et~al., \emph{Phys. Lett.}, \textbf{B223}, 485 (1989).

\bibitem[Benvenuti et~al.(1990)]{bcdmsd}
A.~C. Benvenuti, et~al., \emph{Phys. Lett.}, \textbf{B237}, 592 (1990).

\bibitem[Martin et~al.(2003)]{mrst01e}
A.~D. Martin, R.~G. Roberts, W.~J. Stirling, and R.~S. Thorne, \emph{Eur. Phys.
  J.}, \textbf{C28}, 455--473 (2003), \eprint{hep-ph/0211080}.

\bibitem[Stump et~al.(2003)]{cteq61}
D.~Stump, et~al., \emph{JHEP}, \textbf{10}, 046 (2003),
  \eprint{hep-ph/0303013}.

\bibitem[Alekhin(2003)]{ale02}
S.~Alekhin, \emph{Phys. Rev.}, \textbf{D68}, 014002 (2003),
  \eprint{hep-ph/0211096}.

\end{thebibliography}

\end{document}